\newcommand{\Tr}{\mathop{\mathrm{Tr}}\nolimits}

\documentclass[osajnl,twocolumn,eqsecnum,showpacs]{revtex4}
\usepackage{amssymb,amsbsy,bm,graphicx}

\begin{document}

\title{Comparing omnidirectional reflection from
periodic and quasiperiodic one-dimensional photonic
crystals}

\author{A. G. Barriuso, J. J. Monz\'on,
and L. L. S\'anchez-Soto}
\address{Departamento de \'Optica,
Facultad de F\'{\i}sica,
Universidad Complutense,
28040 Madrid, Spain}

\author{A. Felipe}
\address{Departamento de Estad\'{\i}stica e
Investigaci\'on Operativa I,
Facultad de Matem\'aticas,
Universidad Complutense,
28040 Madrid, Spain}

\bigskip

\begin{abstract}
We determine the range of thicknesses and
refractive indices for which omnidirectional
reflection from quasiperiodic multilayers
occurs. By resorting to the notion of area
under the transmittance curve, we assess
in a systematic way the performance of the
different quasiperiodic Fibonacci multilayers.
\end{abstract}

%\ocis{350.2460 Filters, interference, 230.4170 Multilayers,
%310.6860 Thin films, optical properties}

\maketitle

\section{Introduction}

Photonic crystals are periodically structured
dielectric media possessing photonic band gaps:
ranges of frequency in which strong reflection
occurs for all angles of incidence and all
polarizations. They then behave as
omnidirectional reflectors, free of
dissipative losses. Since the initial predictions
of Yablonovitch~\cite{Yablo87} and John~\cite{John87},
photonic crystals have been attracting a lot
of attention and a wide variety of applications
have been suggested~\cite{Dowling}.

In the one-dimensional case, a photonic crystal
is nothing more than a periodic dielectric
structure. Bragg mirrors consisting of
alternating low- and high-index layers
constitute, perhaps, the archetypical example~\cite{Yeh88}.
In particular, quarter-wave stacks (at normal
incidence) are the most thoroughly studied
in connection with omnidirectional reflection
(ODR)~\cite{Fink98,Dowl98,Yablo98,Chig99,South99,Lekner00}.

The introduction of Fibonacci multilayers by
Kohmoto and coworkers~\cite{KSI87} spurred the
interest for both possible optical
applications~\cite{Schw88} and theoretical
aspects of light transmission in aperiodic
media~\cite{Dulea90,Latge92,Liu97,Vasco98,Macia01}.
In fact, the possibility of obtaining ODR in
quasiperiodic Fibonacci multilayers has been
put forward recently~\cite{Macia98,Cojo01,Lusk01,Peng02,Dong03}.

Underlying all these efforts a crucial
question remains concerning whether
quasiperiodic Fibonacci multilayers would
achieve better performance than usual
periodic ones. To answer such a fundamental
question one first needs to quantify the
idea of ODR performance in a unique manner
that permits unambiguous comparison between
different structures. Only quite recently
a suitable figure of merit has been introduced:
the area under the transmittance curve as a
function of the incidence angle~\cite{Yonte04}.
In this paper, we resort to this concept of
area to rank in a consistent way the ODR
characteristics of these quasiperiodic systems.

\section{Quasiperiodic Fibonacci multilayers}

A Fibonacci system is based on the recursive
relation $S_0 = \{H \}$, $S_1 = \{ L \} $  and
$S_j = S_{j-1} S_{j-2}$ for $j \ge 2$. Here $H$
and $L$ are defined as being two dielectric layers
with refractive indices $(n_H, n_L)$ and thicknesses
$(d_H, d_L)$, respectively. The material $H$ has a
high refractive index while $L$ is of low refractive
index. The number of layers is given by $F_j$,
where $F_j$ is a Fibonacci number obtained from
the recursive law $F_j = F_{j-1} + F_{j-2}$, with
$F_0 = F_1 = 1$.

In order to properly compare the optical response
of these systems we will rely on the transfer-matrix
technique.  The transfer matrix $\mathsf{M}_j$ for the
Fibonacci system $S_j$ can be computed as~\cite{KSI87}
\begin{eqnarray}
& \mathsf{M}_0  = \mathsf{M}_H ,
\qquad
\mathsf{M}_1  = \mathsf{M}_L , &  \nonumber \\
& & \\
& \mathsf{M}_j  =  \mathsf{M}_{j-1}
\mathsf{M}_{j-2}  , \qquad j \ge 2 . &
\nonumber
\end{eqnarray}
The transfer matrix for the single layer $H$ is
\begin{equation}
\mathsf{M}_H  =
\left (
\begin{array}{cc}
\cos \beta_H &  q_H \sin \beta_H \\
\displaystyle
\frac{1}{q_H} \sin \beta_H  & \cos \beta_H
\end{array}
\right ) ,
\end{equation}
and a analogous expression for $L$. Here $\beta_H =
(2 \pi / \lambda) n_H d_H \cos \theta_H$ is the layer
phase thickness, $\theta_H$ being the angle of
refraction, which is determined by Snell law.
The wavelength in vacuum of the incident
radiation  is $\lambda$. The parameter $q_H$
can be written for each basic polarization
($p$ or $s$) as
\begin{equation}
q_H (p) = \frac{n_H \cos \theta}{n \cos \theta_H} ,
\qquad
q_H (s) =  \frac{n \cos \theta}{n_H \cos \theta_H} ,
\end{equation}
where we have assumed that the layer is imbedded in a
medium of refractive index $n$. Henceforth $\theta$
will denote the angle of incidence and, for simplicity,
the surrounding medium we will supposed to be air
($n = 1$).

Let us consider a $N$-period finite structure whose
basic cell is precisely the Fibonacci multilayer $S_j$.
We denote this system as $[S_j]^N$ and its overall transfer
matrix is
\begin{equation}
\mathsf{M}_j^{(N)} = (\mathsf{M}_j)^N .
\end{equation}
When the unit cell is $S_2$, the resulting multilayer
$[S_2]^N$ is periodic. For $j > 2$, $[S_j]^N$ are
quasiperiodic.

The transmittance $\mathcal{T}_j^{(N)}$ is given
in terms of the matrix $\mathsf{M}_j^{(N)}$
as~\cite{KSI87}
\begin{equation}
\mathcal{T}_j^{(N)} =
\frac{4}{|| \mathsf{M}_j^{(N)} ||^2 + 2} ,
\end{equation}
where $|| \mathsf{M}_j^{(N)} ||^2$ denotes the
sum of the squares of the matrix elements.

In the theory of periodic systems it is well
established that band gaps appear whenever
the trace of the basic period satisfies~\cite{Yeh88}
\begin{equation}
\label{ODR}
| \Tr ( \mathsf{M}_j) | \ge 2  .
\end{equation}
This should be worked out for both basic
polarizations. The trace map is a powerful
tool to investigate this condition, especially
when the index $j$ is high~\cite{KKT83}. In
our context, it reads as
\begin{equation}
\Tr ( \mathsf{M}_{j+1}) = \Tr ( \mathsf{M}_j)
\Tr ( \mathsf{M}_{j-1}) - \Tr ( \mathsf{M}_{j-2})  .
\end{equation}
This simple recurrence relation allows us
to compute easily the band gaps. We quote here
the first nontrivial cases, namely, the pure periodic
system $S_2 = \{ LH \}$ and the first quasiperiodic
one $S_3 = \{ LHL \}$, respectively:
\begin{eqnarray}
\label{bands}
& |\cos \beta_L \cos \beta_H -
\Lambda_{LH} \
\sin \beta_L \sin \beta_H | \ge 1 , &
\nonumber \\
& & \\
& |\cos ( 2 \beta_L) \cos \beta_H -
\Lambda_{LH} \
\sin (2 \beta_L) \sin \beta_H | \ge 1 . &
\nonumber
\end {eqnarray}
The function $\Lambda_{LH}$ is
\begin{equation}
\Lambda_{LH} = \frac{1}{2}
\left (
\frac{q_L}{q_H} +
\frac{q_H}{q_L}
\right ) ,
\end {equation}
which is frequency independent but takes
different values for $p$ and $s$ polarizations.
However, one can check that, irrespective of the
angle of incidence, the following relation for both
basic polarizations holds:
\begin{equation}
\label{qps}
\frac{\Lambda_{LH}(p)}{\Lambda_{LH}(s)} \le  1 .
\end{equation}
Due to the restriction (\ref{qps}),
whenever Eqs.~(\ref{bands}) are fulfilled
for $p$ polarization, they are always
true also for $s$ polarization. In
consequence, the $p$-polarization bands
are more stringent than the corresponding
$s$-polarization ones~\cite{Dong03}.

\section{Assessing ODR from quasiperiodic multilayers}

We first investigate the range of layer thicknesses
for which ODR exists; that is, when condition~(\ref{ODR})
holds true for all the incidence angles. Although
for the simple periodic system $S_2$ analytic
approximations are at hand, the general problem
seems to be very involved and we content
ourselves with a numerical exploration.

For definiteness, we fix the refractive indices to
the values $n_L = 1.75$ and $n_H = 3.35$ at
$\lambda = 10 \ \mu$m. In Fig.~1 we have plotted
the zones of ODR for the basic periods $S_j$ (with
$j=2, 3, 4, 5$) in terms of the adimensional thicknesses
$n_L d_L/\lambda$ and $n_H d_H/\lambda$. Note that
the use of these adimensional variables not only
simplifies the presentation of the results, but,
as dispersion  can be neglected, the
results apply to more general situations.

\begin{figure}[h]
\centerline{\includegraphics[height=6cm]{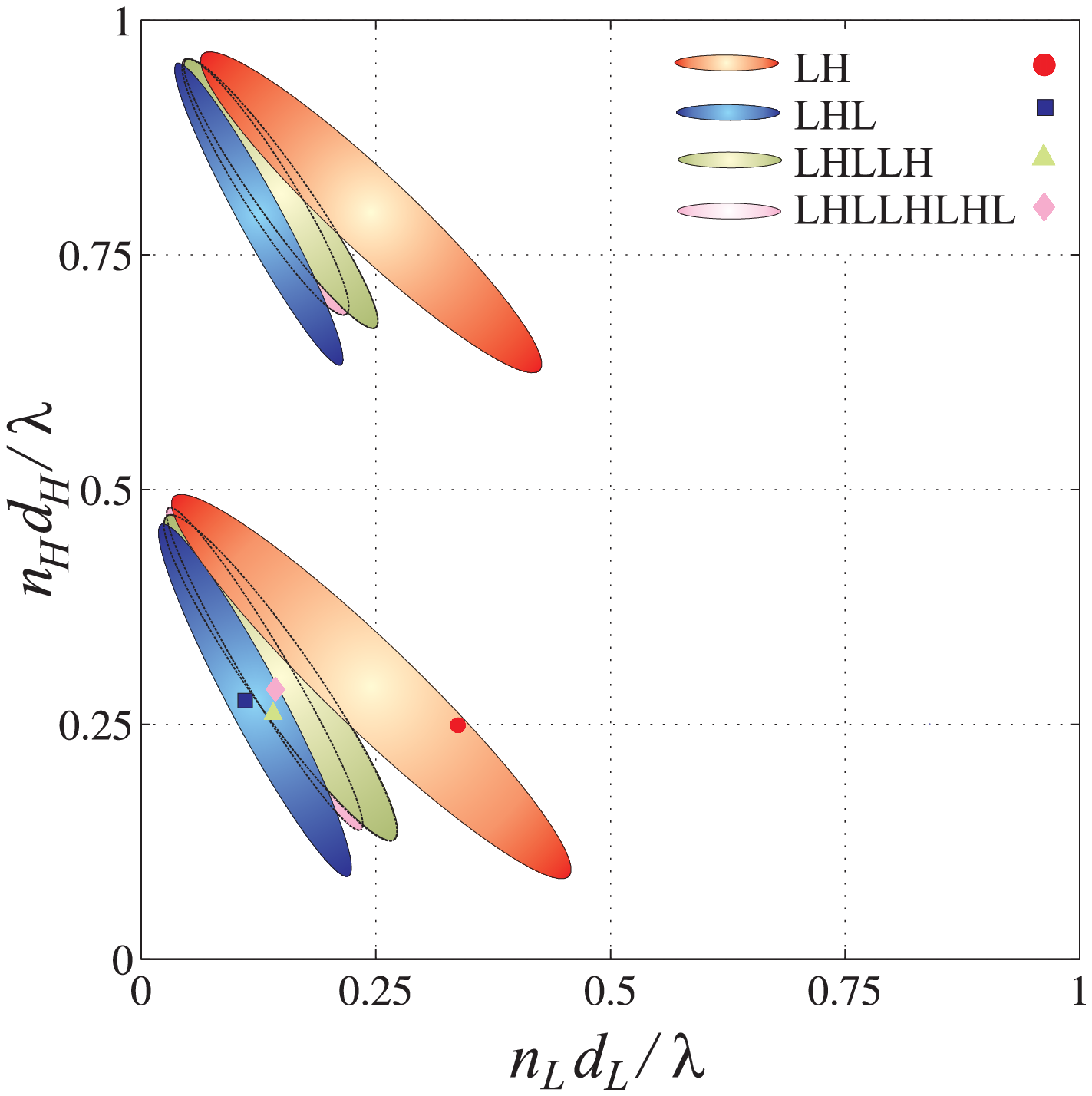}}
\caption{Regions where ODR (for $p$ polarization) occurs
for the Fibonacci systems $S_2 = \{LH\}$, $S_3 = \{LHL\}$,
$S_4 =\{LHLLH\}$, and  $S_5 = \{LHLLHLHL\}$. We have taken
$n_L = 1.75$ and $n_H = 3.35$ at $\lambda = 10 \ \mu$m.
The inset identifies the filled ellipses. The marked
points correspond to the minimum area for each one
of the systems.}
\end{figure}

The contours of all these regions are approximately
elliptical. For every allowed value of $n_L
d_L/\lambda$ there are two values of $n_H d_H/
\lambda$. This can be traced back to the
explicit form of Eqs.~(\ref{bands}) for the band
gaps. The ellipses for $S_2$ are the biggest, which
confirms that this simple system has the best range of
ODR in terms of $n d/\lambda$ variables. Note also
that the usual Bragg solution with layers of a
quarter-wavelength thick at normal incidence,
namely
\begin{equation}
\label{lambda4}
n_L d_L/\lambda = n_H d_H/\lambda = 1/4 ,
\end{equation}
works for $S_2$, but not for the others.

It is worth stressing the fact that the regions of
ODR for $S_2$ and $S_3$ are disjoint. This increases
the difficulty of comparison between these systems.
On the contrary, all the quasiperiodic multilayers
have a significant region of common parameters.
In fact, from the system $S_6$ onwards, all the
elliptic contours are essentially the same as for
the $S_5$.

These regions of ODR are not enough to fully
quantify the performance of the multilayer. In
Ref.~\cite{Yonte04} we have proposed that,
once the materials and the wavelength are fixed,
the area under the transmittance as a function
of the angle of incidence $\theta$
\begin{equation}
\label{area}
\mathcal{A}^{(N)}_j = \int_0^{\pi/2}
\mathcal{T}^{(N)}_j (\theta) \ d \theta ,
\end{equation}
is an appropriate figure of merit for the
structure: the smaller this area, the better
the performance as ODR. In Fig.~2 we have
plotted this area as a function of
$n_L d_L/\lambda$ and $n_H d_H/\lambda$
for $S_2$. The area has been computed solely
for the points fulfilling the ODR condition,
so the abrupt steps give the boundaries of ODR
plotted in Fig.~1.  However, this function
varies significantly in the ODR region.

\begin{figure}
\centering
\centerline{\includegraphics[height=6cm]{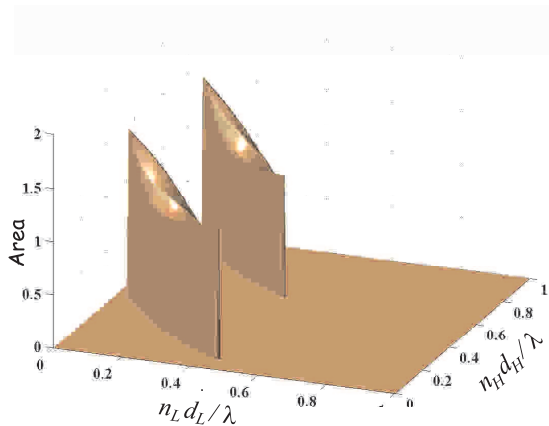}}
%\resizebox{0.75\columnwidth}{!}{\includegraphics{Figure2.eps}}
\caption{Area under the transmittance curve,
defined in Eq.~(\ref{area}), as a function of
$n_L d_L /\lambda$ and $ n_H d_H/ \lambda$ for
the system $S_2$, with the same data as in Fig.~1.}
\end{figure}

In fact, for the present case the minimum of this
area is reached at the point
\begin{equation}
n_L d_L/\lambda = 0.34305  ,
\qquad
n_H d_H/\lambda = 0.25416 .
\end{equation}
While the value of $n_H d_H/ \lambda $ essentially
coincide with the quarter wavelength solution
(\ref{lambda4}),  $n_L d_L/ \lambda $ differs
more than 30 \% of that solution.

\begin{figure}[t]
\centering
\centerline{\includegraphics[height=6cm]{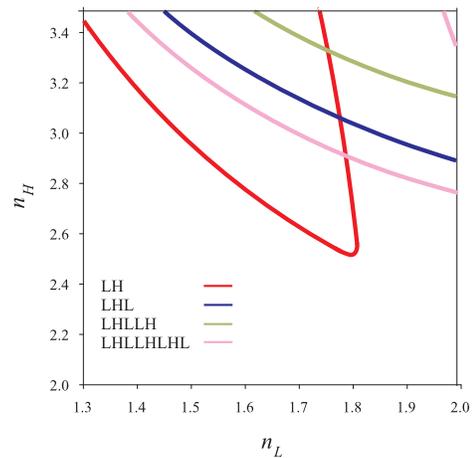}}
%\resizebox{0.75\columnwidth}{!}{\includegraphics{Figure3.eps}}
\caption{Regions of ODR for the same Fibonacci
multilayers as in Fig.~1 in the plane $(n_L, n_H)$
of refractive indices. The curves show the limit
of  ODR for each stack with the optimum thicknesses
marked in Fig.~1.}
\end{figure}

In Fig.~1 we have marked the points of minimum
area for each one of the Fibonacci systems $S_j$.
We see the strong difference for the periodic and
the quasiperiodic systems. In fact, for the latter
($S_j$ with $j \ge 3$) we can summarize the results
saying that the optimum  area is reached approximately
at the values of the parameters
\begin{equation}
n_L d_L/\lambda = 1/8  ,
\qquad
n_H d_H/\lambda = 1/4  .
\end{equation}
In our view, this is a  remarkable result: from the
principle of minimum area, we have consistently derived
optimum parameters for ODR, which differ a lot from
the usual solutions found in the literature.

For the thicknesses giving minimum area, we have
calculated the region in the $(n_L, n_H)$
plane for which ODR occurs. In Fig.~3 we have
plotted the boundary of such a region for
the same Fibonacci multilayers as before: above
such curves we have the ODR region. It is again
the periodic system the first in fulfilling ODR:
the onset of the ODR curve is at $n_H \simeq
2.5$, in agreement with previous estimations~\cite{Lekner00}.

Of course, the optimum parameters for the system
$S_j$ do not need to be optimum for $[S_j]^N$. To
elucidate this question, we have computed numerically
the values of $n_L d_L /\lambda$ and $n_H d_H /\lambda$
for different systems containing up to 42 layers and
for the same indices as before. In Table 1 we have
summarized the corresponding data. We have included
only results for the five first periods $N=1, \ldots, 5$,
since from $[S_j]^5$ onwards, all the thicknesses are
fairly stable, while the area tends rapidly to 0,
as one would expect from a band gap. We can conclude
that the optimum parameters do not depend strongly
on the number of layers.

\begin{figure}[h]
\centerline{\includegraphics[height=6cm]{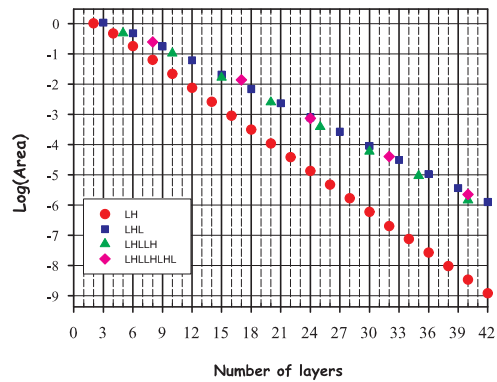}}
%\resizebox{0.85\columnwidth}{!}{\includegraphics{Figure4.eps}}
\caption{Logarithm of the area computed for the systems
$[S_j]^N$ as a function of the number of layers.}
\end{figure}

From previous results for the case of Bragg mirrors,
it is reasonable to assume that the transmittance of
$[S_j]^N$ tends to zero exponentially with the number
of layers. To test such an \textit{ansatz}, we have plotted
the area (in a logarithmic scale) for all these systems.
The results are presented in Fig.~4. We think that a
simple glance at this figure is enough to decide on the
performance of the quasiperiodic systems as omnidirectional
reflectors.

It is quite clear that all the quasiperiodic systems,
irrespective of the index $j$, behave essentially in
the same way as far as ODR is concerned. All the points
for these systems fit into a straight line. On the
other hand, the periodic system $S_2$ lies on another
straight line, but with a better slope. That is,
for a given number of layers of the system (and under
the hypothesis of optimum thicknesses), the system
$[LH]^N$ offers better performance than any other.

\begin{table}
\caption{Optimum parameters for the systems $[S_j]^N$
for different basic periods $S_j$ and the number of
periods $N$ ranging from 1 to 5.}
\begin{center}
\begin{tabular}{cccccc}
\textbf{Basic} & \textbf{Bandwidth} & $\# $ \textbf{Periods} &
$\mathbf{n_L d_L/\lambda}$ & $\mathbf{n_H d_H/\lambda}$ &
\textbf{Area}  \\
\hline
\ & \            & 1 & 0.34305 & 0.25416 & 1.01660 \\
\ & \            & 2 & 0.33807 & 0.22187 & 0.47147 \\
$S_2$ & 0.217    & 3 & 0.30817 & 0.23429 & 0.17894 \\
\ & \            & 4 & 0.29821 & 0.23926 & 0.06294 \\
\ & \            & 5 & 0.29572 & 0.24422 & 0.02171 \\
\hline
\ & \            & 1 & 0.11978 & 0.26906 &   1.07725 \\
\ & \            & 2 & 0.12917 & 0.26657 &   0.47955 \\
$S_3$ & 0.233    & 3 & 0.13319 & 0.26409 &   0.17693 \\
\ & \            & 4 & 0.13587 & 0.26409 &   0.06092  \\
\ & \            & 5 & 0.13722 & 0.26409 &   0.02055  \\
\hline
\ & \            & 1 & 0.14795 & 0.25912 & 0.48632 \\
\ & \            & 2 & 0.16538 & 0.25912 & 0.10423 \\
$S_4$ & 0.195    & 3 & 0.16002 & 0.29389 & 0.01647 \\
\ & \            & 4 & 0.16136 & 0.29389 & 0.00251 \\
\ & \            & 5 & 0.16270 & 0.29389 & 0.00038 \\
\hline
\ & \            & 1 & 0.14929 & 0.28396 &  0.24841 \\
\ & \            & 2 & 0.15063 & 0.28396 &  0.01365 \\
$S_5$ & 0.198    & 3 & 0.15331 & 0.28396 &  0.00073 \\
\ & \            & 4 & 0.15197 & 0.28644 &  0.00004 \\
\ & \            & 5 & 0.15331 & 0.28644 &  0.00002 \\
\hline
\end{tabular}
\end{center}
\end{table}

Of course, one may think that the bandwidth of these
systems is different. Sometimes the bandwidth is
defined at normal incidence, and then it has been
argued that quasiperiodic systems offer fundamental
advantages~\cite{Lusk01}.  If we denote by
$\lambda_{\mathrm{short}}$ and $\lambda_{\mathrm{long}}$
the longer- and shorter-wavelength edges for given
ODR bands (of the basic period), it seems more
appropriate to define the ODR bandwidth as~\cite{South99}
\begin{equation}
B = \frac{\lambda_{\mathrm{long}} -
\lambda_{\mathrm{short}}}{\frac{1}{2}
(\lambda_{\mathrm{long}} +
\lambda_{\mathrm{short}})} .
\end{equation}
Note that this is the appropriate definition in our case.
Obviously, the parameters chosen for the purpose of
comparison must be the ones giving minimum area; i. e.,
optimum ODR behavior. In fact, we have numerically
checked that the parameters giving optimum area offer
also a good bandwidth. In Table 1 we have indicated
this parameter, confirming again that with the proper
definition the periodic system is the best.

\section{Concluding remarks}

In summary, we have exploited the idea of minimum area
to fully assess in a systematic way the performance of
omnidirectional reflectors. Although quasiperiodic
systems has attracted a lot of interest due to their
unusual physical properties, Bragg reflectors offer
the best performance, although not at a quarter-wavelength
thick at normal incidence. We believe that the best
feature of our approach is that it provides  a very
clear thread to deal with omnidirectional reflection
properties in a systematic way. Our method is general
and can be applied to any spectral region.


\begin{references}
\bibitem{Yablo87}
E. Yablonovitch, ``Inhibited spontaneous emission in solid-state
physics and electronics," Phys. Rev. Lett. \textbf{58}, 2059-62
(1987).

\bibitem{John87}
S. John, ``Strong localization of photons in certain disordered
dielectric superlattices," Phys. Rev. Lett. \textbf{58}, 2486-9
(1987).

\bibitem{Dowling}
A complete and up-to-date bibliography on the subject can be found
at
\url{http://home.earthlink.net/\symbol{126}jpdowling/pbgbib.html}

\bibitem{Yeh88}
P. Yeh, \textit{Optical Waves in Layered Media} (Wiley, New York,
1988)

\bibitem{Fink98}
Y. Fink,  J. N. Winn, S. Fan, C. Chen, J. Michel, J. D.
Joannopoulos, and E. L. Thomas, ``A dielectric omnidirectional
reflector," Science \textbf{282}, 1679-82 (1998).

\bibitem{Dowl98}
J. P. Dowling, ``Mirror on the wall: you're omnidirectional after
all?," Science \textbf{282}, 1841-2 (1998).

\bibitem{Yablo98}
E. Yablonovitch, ``Engineered omnidirectional
external-reflectivity spectra from one-dimensional layered
interference filters," Opt. Lett. \textbf{23}, 1648-9 (1998).

\bibitem{Chig99}
D. N. Chigrin, A. V. Lavrinenko, D. A. Yarotsky, and  S. V.
Gaponenko, ``Observation of total omnidirectional reflection from
a one-dimensional dielectric lattice," Appl. Phys. A \textbf{68},
25-8 (1999).

\bibitem{South99}
W. H. Southwell, ``Omnidirectional mirror design with quarter-wave
dielectric stacks," Appl. Opt. \textbf{38}, 5464-7 (1999).

\bibitem{Lekner00}
J. Lekner ``Omnidirectional reflection by multilayer dielectric
mirrors," J. Opt. A \textbf{2}, 349-53 (2000).

\bibitem{KSI87}
M. Kohmoto, B. Sutherland, and K. Iguchi, ``Localization of
optics: Quasiperiodic media," Phys. Rev. Lett. \textbf{58}, 2436-8
(1987).

\bibitem{Schw88}
C. Schwartz, ``Reflection properties of pseudorandom multilayers,"
Appl. Opt. \textbf{27}, 1232-4 (1988).

\bibitem{Dulea90}
M. Dulea, M. Severin, and R. Riklund, ``Transmission of light
through deterministic aperiodic non-Fibonaccian multilayers,"
Phys. Rev. B \textbf{42}, 3680-9 (1990).

\bibitem{Latge92}
A. Latg\'e and F. Claro, ``Optical propagation in multilayered
systems," Opt. Commun. \textbf{94}, 389-96 (1992).

\bibitem{Liu97}
N. H. Liu, ``Propagation of light waves in Thue-Morse dielectric
multilayers," Phys. Rev. B \textbf{55}, 3543-7 (1997).

\bibitem{Vasco98}
M. S. Vasconcelos and E. L. Albuquerque, ``Transmission
fingerprints in quasiperiodic dielectric multilayers," Phys. Rev.
B \textbf{59}, 11128-31 (1999).

\bibitem{Macia01}
E. Maci\'a, ``Exploiting quasiperiodic order in the design of
optical devices," Phys. Rev. B \textbf{63}, 205421 (2001).

\bibitem{Macia98}
E. Maci\'a, ``Optical engineering with Fibonacci dielectric
multilayers," Appl. Phys. Lett. \textbf{73}, 3330-2 (1998).

\bibitem{Cojo01}
E. Cojocaru, ``Forbidden gaps in finite periodic and
quasi-periodic Cantor-like dielectric multilayers at normal
incidence," Appl. Opt. \textbf{40} 6319-26 (2001).

\bibitem{Lusk01}
D. Lusk, I. Abdulhalim and F. Placido, ``Omnidirectional
reflection from Fibonacci quasi-periodic one-dimensional photonic
crystal," Opt. Commun. \textbf{198}, 273-9 (2001).

\bibitem{Peng02}
R. W. Peng, X. Q. Huang, F. Qiu, M. Wang, A. Hu, S. S. Jiang, and
M. Mazzer, ``Symmetry-induced perfect transmission of light waves
in quasiperiodic dielectric multilayers," Appl. Phys. Lett.
\textbf{80}, 3063-5 (2002).

\bibitem{Dong03}
J. W. Dong, P. Han, and H. Z. Wang, ``Broad omnidirectional
reflection band forming using the combination of Fibonacci
quasi-periodic and periodic one-dimensional photonic crystals."
Chin. Phys. Lett. \textbf{20}, 1963-5 (2003).

\bibitem{Yonte04}
T. Yonte, J. J. Monz\'on, A. Felipe, and L. L. S\'anchez-Soto,
``Optimizing omnidirectional reflection by multilayer mirrors," J.
Opt. A \textbf{6}, 127-31 (2004).

\bibitem{KKT83}
M. Kohmoto, L. P. Kadanoff, and C. Tang, ``Localization Problem in
One Dimension: Mapping and Escape," Phys. Rev. Lett. \textbf{50},
1870-2 (1983).

\bibitem{Lek87}
J. Lekner, \textit{Theory of Reflection} (Dordrecht, The
Netherlands, 1987).
\end{references}
\end{document}